\documentclass[aps,twocolumn,superscriptaddress]{revtex4-2}

\usepackage{graphicx}      
\usepackage{dcolumn}       
\usepackage{amsmath}       
\usepackage{amssymb}       
\usepackage{xcolor}        
\usepackage{hyperref}      
\usepackage{natbib} 

\usepackage{bm}            
\usepackage{booktabs}      
\usepackage{multirow}      
\usepackage[T1]{fontenc}
\usepackage{lmodern}
 \usepackage{microtype}
\usepackage{subcaption}
\usepackage{caption}
\makeatletter

\renewcommand\footnote[1]{%
  \stepcounter{footnote}%
  \protected@xdef\@thefnmark{\thefootnote}%
  \@footnotemark\@footnotetext{#1}%
}
\makeatother

\begin{document}

\title{A New Method for Quasinormal Modes From Bound States and Homotopy deformations}

\author{Hao-Yun Ma}
\affiliation{Key Laboratory of Atomic and Subatomic Structure and Quantum Control (Ministry of Education), Guangdong Basic Research Center of Excellence for Structure and Fundamental Interactions of Matter, School of Physics, South China Normal University, Guangzhou 510006, China and Guangdong Provincial Key Laboratory of Quantum Engineering and Quantum Materials, Guangdong-Hong Kong Joint Laboratory of Quantum Matter, South China Normal University, Guangzhou 510006, China}
\author{Bo-Yun Zhou}
\affiliation{Key Laboratory of Atomic and Subatomic Structure and Quantum Control (Ministry of Education), Guangdong Basic Research Center of Excellence for Structure and Fundamental Interactions of Matter, School of Physics, South China Normal University, Guangzhou 510006, China and Guangdong Provincial Key Laboratory of Quantum Engineering and Quantum Materials, Guangdong-Hong Kong Joint Laboratory of Quantum Matter, South China Normal University, Guangzhou 510006, China}
\author{Jia-Hui Huang}
\email{huangjh@m.scnu.edu.cn} 
\affiliation{Key Laboratory of Atomic and Subatomic Structure and Quantum Control (Ministry of Education), Guangdong Basic Research Center of Excellence for Structure and Fundamental Interactions of Matter, School of Physics, South China Normal University, Guangzhou 510006, China and Guangdong Provincial Key Laboratory of Quantum Engineering and Quantum Materials, Guangdong-Hong Kong Joint Laboratory of Quantum Matter, South China Normal University, Guangzhou 510006, China}


\begin{abstract}
Inspired by Mashhoon's bound state method, we propose a new bound state method for computing quasinormal modes (QNMs). By a two-step coordinate transformation where a real parameter $\alpha$ is introduced, a QNM problem is mapped to a bound state problem, whose eigenvalues $E_n(\alpha)$ are inversely mapped to the QNM frequencies $\omega_n$ via analytic continuation. With this method, we numerically calculate various QNM frequencies for a Schwarzschild black hole directly from the bound state spectrum of the inverted Regge-Wheeler potential for the first time. It is found that the method yields QNM frequencies of high accuracy for low-lying modes with overtone $n\leq\ell$ ($\ell$ is the multipole number), while the accuracy degrades or the calculation fails for higher overtones. To identify the origin of this limitation, we analyze the singularity structure of the eigenvalues $E_n(\alpha)$ using Padé approximants in the complex $\alpha$-plane. For higher overtones, the singularities of $E_n(\alpha)$ lie within the analytic continuation circle, providing a direct explanation for the limitation of the method. To mitigate this limitation, we suggest a homotopy deformation to the potential, which improves the method and enable us to compute a few more high overtone modes reliably.  
\end{abstract}

\maketitle


\textit{Introduction---}The black hole quasinormal modes (QNMs) are characteristic oscillations of a perturbed black hole, which encode the geometric information of the black hole \cite{Kokkotas:1999bd,Riview09,Konoplya:2011qq}. 
The QNMs are expected to play key roles in interpreting the ringdown signals of binary black hole mergers in gravitational-wave astronomy~\cite{Berti:2005ys,Berti:2007zu,Baibhav:2023clw,Riview2026},  testing general relativity \cite{Dreyer2004} and its central predictions of general relativity, such as the no-hair theorem \cite{Isi:2019aib} and the event horizon of a black hole \cite{ Cardoso:2016rao}, and studying the matter distribution around a black hole \cite{Leung:1997was,Barausse:2014tra,Cardoso:2024mrw,Yang:2024vor,Ianniccari:2024ysv,Laeuger:2025zgb} and modified theory of general relativity \cite{Berti:2015itd,Maselli:2023khq,Volkel:2022khh,McManus:2019ulj,Cardoso:2019mqo,Cano:2024ezp,Cano:2024jkd}. Besides these important applications, 
 the QNMs also has purely theoretical significance and can be used to investigate the linear stability of black holes \cite{Brito:2015oca}.


Given the extensive studies on QNMs, many computational methods have been developed over the years. 
One of the well-established numerical methods is Leaver's continued fraction method, which achieves exceptionally high numerical precision. 
However, it relies on the derivation of a three-term recurrence relation and provides no analytical intuition for QNMs ~\cite{Leaver1985,Leaver1986,Leaver1990}. 
In contrast, the Wentzel-Kramers-Brillouin (WKB) method can yield approximate analytical results for the QNMs, but loses accuracy when computing low multipole or high overtone QNMs. To mitigate the accuracy, many researchers have been continuously refining this method~\cite{Schutz1985,Iyer1987I,Iyer1987II,Kokkotas1988,Seidel1990,Guinn1990,Kokkotas1993,Andersson1993,Matyjasek2017,Konoplya2019}.

In the 1980s, Mashhoon et al. put forward an elegant analytical method ~\cite{Mashhoon1982,Ferrari1984,Blome1984,Mashhoon1984}, which relates the QNM frequencies of a potential barrier to the bound state spectra of the corresponding inverted potential well through an analytic continuation of the relevant parameters. 
To be concrete, suppose the radial master equation for perturbations of a spherically symmetric black hole can be written in tortoise coordinate $x$ as
\begin{equation}
\frac{d^2}{dx^2}\psi(x) + \left[\omega^2 - V(x,P)\right]\psi(x) = 0,
\label{eq:master}
\end{equation}
where $\omega$ is the QNM frequency, $V(x,P)$ is a real effective potential and $P$ is a set of model parameters.
For the Schwarzschild black hole (SchBH),  the effective potential for scalar ($s=0$), electromagnetic ($s=1$), and odd-parity gravitational ($s=2$) perturbations is the Regge-Wheeler (RW) potential~\cite{Regge:1957td,Teukolsky1973} 
\begin{equation}\label{eq:RW}
	V(r,P) = \left(1- \frac{2M}{r}\right)  \left(\frac{\ell(\ell+1)}{r^2}+\frac{2M(1-s^2)}{r^3}\right),
\end{equation}
and the tortoise coordinate is defined as  $x = r + 2M\ln\left(r/2M - 1\right)$. 

Mashhoon's insight~\cite{Mashhoon1984} was that a coordinate transformation $x \to -ix$ maps the QNM problem to a bound state problem. Under this transformation, Eq.~\eqref{eq:master} becomes
\begin{equation}
\frac{d^2}{dx^2}\Psi(x) + \left[E - V_{inv}(-ix,P)\right]\Psi(x) = 0,
\label{eq:complex}
\end{equation}
where $E\equiv-\omega^2$, the inverted potential is $V_{inv}(-ix,P)=-V(-ix,P)$. If there is a mapping $P\to\pi(P)$ satisfying
\begin{equation}\label{potential trans}
	V(-ix,\pi(P))=V(x,P),
\end{equation}
then, Eq.\eqref{eq:complex} becomes a standard bound state eigenvalue problem after the mapping.
The QNM frequency with overtone $n$ can be obtained from the bound state eigenvalue $E_n(P)$ via an inverse transformation $\omega_n = \sqrt{-E_n(\pi^{-1}(P))}$ involving analytic continuation of certain parameters.  
Although conceptually appealing, a major limitation on this method is that it needs closed-form analytical bound state spectra for the potential. 
Thus, this method is typically applied to the Pöschl-Teller (PT)~\cite{Pöschl1933} and Eckart~\cite{Eckart1930} potentials, which are used to locally approximate the effective potential of a black hole near its extreme. 
Extensions of the application can also be found in~\cite{Zaslavskii1991,Galtsov1992,Bender1973,Sulejmanpasic2018,Hatsuda:2019eoj,Matyjasek2019,Aminov2020,Richarte2024}. 

A few years ago, a significant step to circumvent the limitation was taken by Völkel, who proposed computing the bound state energies numerically and constructing a Taylor expansion in model parameters for analytic continuation~\cite{Volkel2022,Volkel2025}. Völkel validated this approach for the analytically solvable PT potential, and further demonstrated it for the Breit-Wigner (BW) and mixed (PT+BW) potentials, which have no analytical bound state spectra.

In this work, we propose a new numerical method that enables us to calculate QNM frequencies from bound state eigenvalues by an analytic continuation in a single parameter. 
As an application, we numerically calculate the scalar, electromagnetic and gravitational QNM frequencies of SchBH from the exact RW potential. 
To identify the inherent limitation of this method, we numerically analyzed the analytic structure of the bound state spectra.
This analysis reveal the analytic obstacle causing the breakdown of this method at high overtones. To mitigate this situation, 
we further introduce a homotopy deformation to the inverted potential, which changes the analytic structure of bound state spectra and improve the accuracy of the high overtone results. Throughout this work, we use units where $G = c = 1$.


\textit{Method---}Start with Eq.~\eqref{eq:complex}, the traditional method needs a parameter transformation $P\to\pi(P)$ satisfying Eq.~\eqref{potential trans} and depending on the specific form of the potential. In this work, we do not consider a transformation of the model parameters, but instead introduce a real scaling parameter $\alpha$ into the potential in Eq.~\eqref{eq:complex} via a substitution $-ix\to\alpha x$. For any real value $\alpha\neq 0$, Eq.~\eqref{eq:complex} becomes
\begin{equation}
\frac{d^2}{dx^2}\Psi(x) + \left[E - V_{inv}(\alpha x,P)\right]\Psi(x) = 0,
\label{eq:bound}
\end{equation}
where $V_{inv}(\alpha x,P)=-V(\alpha x,P)$. This equation defines a standard bound state eigenvalue problem whose energies $E_n(\alpha)$ can be computed numerically.  At the target point $\alpha=-i$, Eq.~\eqref{eq:bound} becomes exactly to the original QNM problem in Eq.~\eqref{eq:complex}, so the QNM frequencies can be obtained via analytic continuation of the bound state eigenvalues $E_n(\alpha)$, i.e., $\omega_n = \sqrt{-E_n(\alpha=-i)}$. 

In practice, we numerically compute $\{E_n(\alpha_j)\}$ on a uniform grid of real values $\alpha$ centered at $\alpha_0$ with step-size $h$ using the shooting method, fit the data to a numerical representation of $E_n(\alpha)$ using Taylor expansions and Padé approximants, and then perform the analytic continuation of $E_n(\alpha)$ to obtain QNM frequencies. Since the shooting method and Padé approximants are well-established techniques, we refer the reader to Appendices~\ref{sec:Appendix A} and~\ref{sec:Appendix B} for more details.


\textit{Results---}We apply the new numerical method to calculate the scalar, electromagnetic, and axial gravitational perturbations of the SchBH. All results in this section are obtained using a 31-point finite-difference stencil with step-size $h=0.02$ centered at $\alpha_0=0.5$, and the bound state eigenvalues are computed to a precision of more than 35 decimal places. In the main text, we focus on axial gravitational perturbations with $\ell=2$. The QNM frequencies calculated with the new bound state method via Taylor expansions (30th-order) and Padé approximants are listed in Table~\ref{tab:QNM_Taylor+Pade}, and the relative errors of their real and imagine parts with respect to reference values from Leaver's continued fraction method~\cite{Leaver1985,Leaver1986,Leaver1990,dataBerti} are given in Table~\ref{tab:QNM_Taylor+Pade err}. 
Results for scalar and electromagnetic perturbations via Padé approximants are provided in Appendix~\ref{sec:Appendix D} (Tables~\ref{tab:QNM_scalar} and~\ref{tab:QNM_EM}).

\begin{table*}[htbp]  
\caption{QNM frequencies for gravitational perturbations ($\ell=2$, $n\in[0,4]$) of SchBH. }
  \label{tab:QNM_Taylor+Pade}
  \setlength{\tabcolsep}{17.5pt}  
  \begin{tabular}{c c c c}

    \toprule
    $n$& $\omega_{\rm Taylor}$&$\omega_{\rm \text{Pad\'e}}$&$\omega_{\rm Leaver}$ \\
    \midrule
     $0$&  $0.3736838989+0.0889544568i$& $0.3736716847+0.0889623155i$
&  $0.3736716844+0.0889623157i$\\
    $1$&  $0.3467672354+0.2738596227i$
&$0.3467109504+0.2739148839i$&  $0.3467109969+0.2739148753i$\\
     $2$&  $0.3033906872+0.4781977762i$
& $0.3010560088+0.4782760170i$
&  $0.3010534546+0.4782769832i$\\
     $3$& $2.0817536316 - 2.3943681414i$& $0.2498769044+0.7075502628i$
&  $0.2515049622+0.7051482024i$\\
     $4$&$10.001317665+74.289764364i$& $0.2170024620+0.9323534842i$
& $0.2075145798+0.9468448909i$\\

    \bottomrule
  \end{tabular}
\end{table*}

\begin{table}[htbp]  
  \caption{
  Relative error of QNM frequencies for gravitational perturbations with $\ell=2, n\in[0,4]$.
  }
  \label{tab:QNM_Taylor+Pade err}
\setlength{\tabcolsep}{5pt}  
  \begin{tabular}{c  c c c c}

    \toprule
    $n$& $\delta_{\rm Re(Taylor)}$& $\delta_{\rm Im(Taylor)}$& $\delta_{\rm\text{Re(Pad\'e)})}$&$\delta_{\rm\text{Im(Pad\'e)}}$ \\
    \midrule
     $0$&  $3.3\times 10^{-05}$&$8.8\times 10^{-05}$& $7.5\times 10^{-10}$&$2.1\times 10^{-09}$
\\
    $1$&  $1.6\times 10^{-04}$&$2.0\times 10^{-04}$ 
&$1.3\times 10^{-07}$&$3.1\times 10^{-08}$\\
     $2$& $7.8\times 10^{-03}$&$1.7\times 10^{-04}$&$8.5\times 10^{-06}$&$2.0\times 10^{-06}$\\
     $3$& $-$&$-$& $6.5\times 10^{-03}$&$3.4\times 10^{-03}$\\
     $4$&$-$&$-$& $4.6\times 10^{-02}$&$1.5\times 10^{-02}$\\
    \bottomrule
  \end{tabular}
\end{table}
As presented in Table~\ref{tab:QNM_Taylor+Pade err}, the accuracy of the results with Taylor expansion gradually declines as $n$ increases and a breakdown emerges at $n=3$. We check that the results with Taylor expansions converge for $n=0,1,2$ but diverge for $n\geq3$.  The Padé approximants yield higher accuracy across all QNM frequencies; nevertheless, their accuracy also degrades with increasing $n$. The divergence in the Taylor-expansion case indicates that for $n\geq3$ there is an intrinsic obstruction for the analytic continuation of $E_n(\alpha)$ from $\alpha_0$ to $\alpha=-i$.
To identify this obstruction, we numerically examine the analytic structure of $E_n(\alpha)$ in the complex $\alpha$-plane.

\textit{The analytic structure of $E_n(\alpha)$ from Padé analysis---}For analytically solvable potentials such as the PT potential, the singularities of $E_n(\alpha)$ can be determined directly from its analytical expression \cite{Li:2026ptw}. However, since the bound state spectrum of the exact RW potential has no closed-form analytical expression, the singularities of $E_n(\alpha)$ can not be analyzed directly. We therefore examine its singularities numerically by analyzing the pole distributions of its Padé approximants $R_{[N/M]}(\alpha)=P_N(\alpha)/Q_M(\alpha)$. 

As discussed in Ref.~\cite{PhysRevD.106.114022}, the Nuttall–Pommerenke theorem~\cite{POMMERENKE1973775} guarantees that for a meromorphic function, the diagonal Padé sequence $[M/M]$ converges to it in logarithmic capacity in any compact set of the complex plane. Stable poles for sufficiently large $M$ in these sequences can be identified as  
poles of the underlying meromorphic function. In finite-order Padé approximants, Froissart doublets~\cite{1975Essentials,masjuan2010,Yamada2014} are ubiquitous and 
spurious poles that depend strongly on the degree of the polynomial and should be omitted in the identification. 
 In this work, we adopt the diagonal Padé sequence $[M/M]$ with $M\in[21,30]$, constructed from a 61-point data set $\{E_n(\alpha_j)\}$ with step-size $h=0.01$ centered at $\alpha=0.5$. In practice, for each diagonal Padé order, we exclude poles whose nearest zeros lie within $|\alpha_{pole}-\alpha_{zero}|<0.001$. 
 Poles that shift substantially from one order to the other or appear only at isolated orders, are also regarded as false poles. 
The poles that locate at nearly the same positions and appear across those different Padé orders are identified as the approximate singularities of $E_n(\alpha)$.

\begin{figure}[ht]
    \centering
    \includegraphics[width=0.95\linewidth]{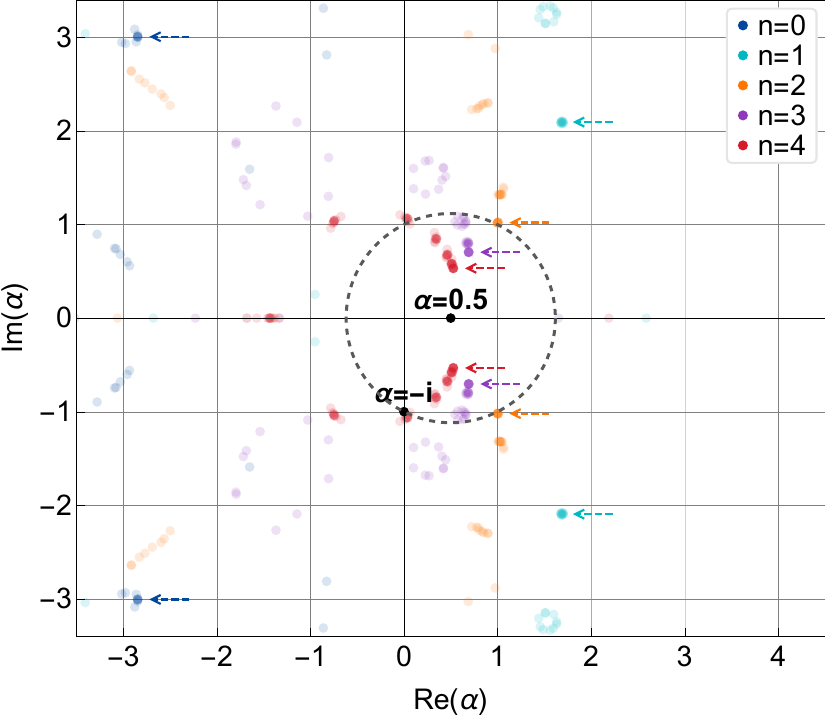}
    \caption{Pole distributions of the diagonal Padé appro{-\allowbreak}ximants for each $E_n(\alpha
   )$$(\ell=2)$. Semi-transparent dots denote poles at individual Pad\'e orders. Overlapping poles at nearly the same positions form darker spots; the darkest spots (marked by arrows), which arise from poles that overlap across the widest range of orders, indicate the approximate positions of singularities of $E_n(\alpha
   )$. The dashed circle centered at $\alpha=0.5$ marks the boundary of analytic continuation.}
   \label{fig:H=1}
\end{figure}

Fig.~\ref{fig:H=1} shows the pole distributions of the diagonal Padé approximants of $E_n(\alpha) (n\in[0,4])$ for gravitational perturbations with $\ell=2$. The positions of darkest spots for each eigenvalue are identified as the approximate singularities of $E_n(\alpha)$. The positions of the singularities depend remarkably on the overtone number $n$: as $n$ increases, the singularities lie closer to the expansion center $\alpha_0=0.5$. As a result, for $n=0,1,2$, the nearest singularities to $\alpha_0$ lie outside or nearly on the required continuation circle, and thus the Taylor expansion converges; for $n=3$ and $n=4$, the corresponding singularities lie inside the circle, which directly explains why the Taylor expansion diverges for these modes. 
Since the radius of the analytical continuation is determined by the distance from the expansion center $\alpha_0$ to the target point $\alpha=-i$, one can infer that a smaller value of $\alpha_0$ is preferred in the calculation to avoid including singularities of $E_n(\alpha)$.


 \textit{A homotopy deformation of the inverted potential---}The pole analysis reveals that the failure to obtain higher overtone QNM frequencies with Taylor expansion is due to the fact that singularities of $E_n(\alpha)$ lie within the radius of analytic continuation. This suggests that modifying the analytic structure of $E_n(\alpha)$ may enable us to calculate higher overtone modes. To achieve this, we suggest performing a homotopy deformation to the inverted potential \cite{Eugene1990}
 \begin{equation}
V_{\rm \lambda}(\alpha x)=\mathcal{H}(\alpha,\lambda) V_{ inv}(\alpha x),
\label{eq:deepening potential}
\end{equation}
 where the scaling factor is defined as
 \begin{equation}
\mathcal{H}(\alpha,\lambda  ) = (\lambda  - 1)\alpha^2 +\lambda ,
\label{eq:deepening}
\end{equation}
which equals to 1 when $\alpha=-i$.
\begin{figure}[b]
    \centering
    \includegraphics[width=0.95\linewidth]{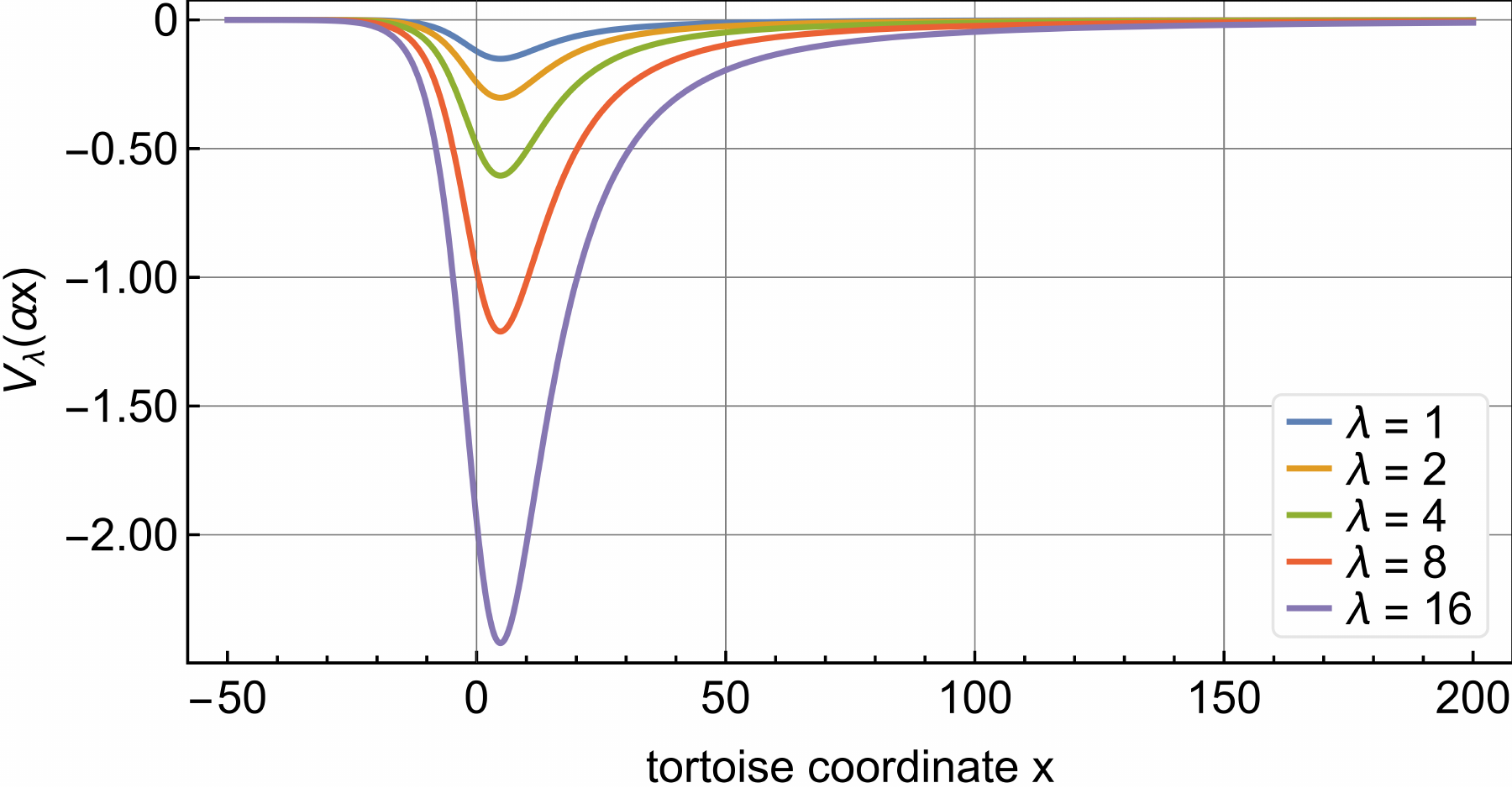}
    \caption{The homotopy deformed inverted RW potential for gravitational pertur{-\allowbreak}bations with $\ell=2$ and $\alpha=0.5$. }
   \label{fig:potential_H}
\end{figure}

Fig.~\ref{fig:potential_H} shows the potential $V_\lambda(\alpha x)$ with different values of $\lambda$ with $\alpha=0.5$. When $\lambda=1$, it reduces to the original inverted RW potential $V_{ inv}(\alpha x)$. The parameter $\lambda$ serves as a homotopy parameter, and as $\lambda$ increases, the potential well is continuously deepened by $\mathcal{H}(\alpha,\lambda)$ for real $\alpha$. 
Since $\mathcal{H}(\alpha=-i,\lambda) = 1$ holds for all $\lambda$, the black hole QNM problem is exactly restored at the target point $\alpha=-i$. Thus, we can calculate the bound state eigenvalues for a deeper potential well with a larger $\lambda$, and then obtain the QNM frequencies via the same analytic continuation with $\alpha=-i$. 

\begin{figure}[t]  
\centering
\begin{subfigure}[b]{0.48\columnwidth}  
\centering
        \includegraphics[width=1\textwidth]{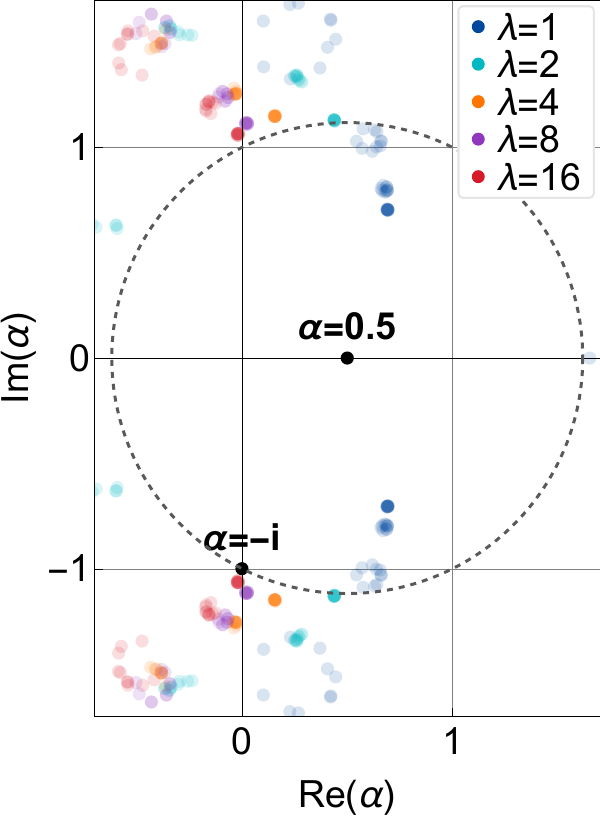}
        \caption{$n=3$}
        \label{fig:pole_n3}
    \end{subfigure}
    \hfill
    \begin{subfigure}[b]{0.48\columnwidth}
        \centering
        \includegraphics[width=1\textwidth]{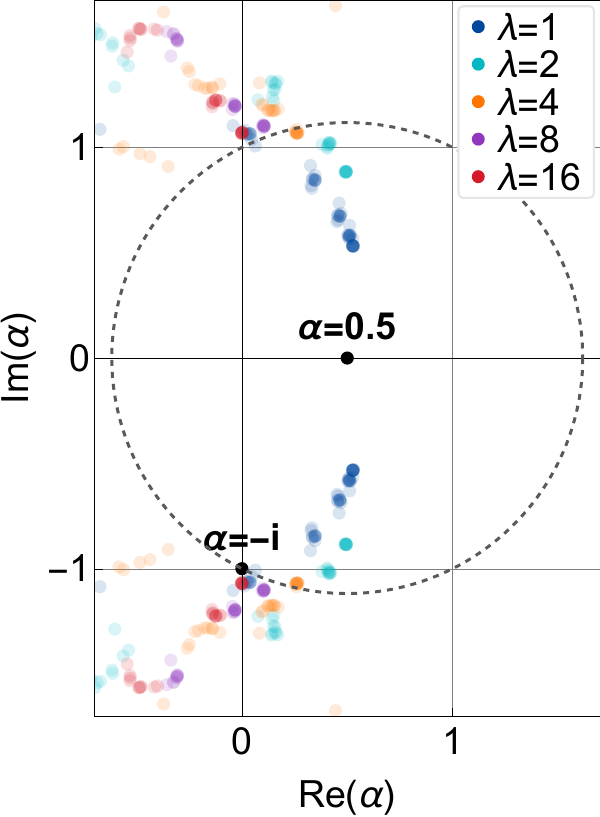}
        \caption{$n=4$}
        \label{fig:pole_n4}
    \end{subfigure}
 \caption{Poles of the diagonal Pad\'e approximants for $E_n(\alpha)$ ($\ell=2, n=3,4$) with different values of $\lambda$. The interpretation of lighter and darker points follows Fig.~\ref{fig:H=1}.}
    \label{fig:pole_migration}
\end{figure}

Then, we investigate how the homotopy deformation affects the distribution of singularities of $E_n(\alpha)$ and the convergence properties of the corresponding QNM frequencies. For illustration, we focus on the $n=3$ and $n=4$ cases since their poles lie inside the analytic continuation circle for the original inverted RW potential. Fig.~\ref{fig:pole_migration} 
shows the pole distributions of the diagonal Padé approximants for these modes with different values of $\lambda$. A clear pattern emerges: as $\lambda$ increases, the poles shift outward and eventually cross the boundary of the analytic continuation circle; however, beyond a certain threshold, further increases of $\lambda$ yield no additional outward shift;
another interesting observation is the poles gradually approach the target point $\alpha=-i$ as $\lambda$ increases.   

\begin{figure}[htbp]
    \centering
\includegraphics[width=0.9\linewidth]{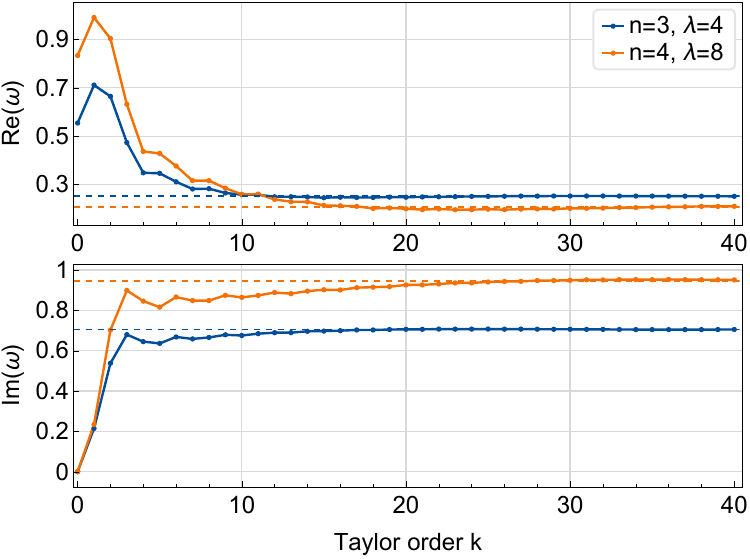}
    \caption{
Convergence curves of the QNM frequencies ($\ell= 2, n
=3,4$) with increasing Taylor order with specific $\lambda$. Dashed lines represent reference values from Leaver’s method.}
    \label{fig:Taylor_H}
\end{figure}

The above homotopy deformation to the original inverted potential directly remedies the previous failure to obtain the $n=3$ and $n=4$ overtones in Table~\ref{tab:QNM_Taylor+Pade} from Taylor expansions. 
Fig.~\ref{fig:Taylor_H} shows the convergence behavior of the Taylor-expansion results for the $n=3$ and $n=4$ QNM frequencies with $\lambda=4$ and $\lambda=8$, respectively.
These $\lambda$ values are chosen such that the corresponding poles are shifted outside the analytic continuation circle. 
Explicitly, the numerical values of the two QNM frequencies obtained from a 30th-order Taylor expansion are $\omega_3=0.2524800930 + 0.7060299311 i$ and $\omega_4= 0.2019118777 + 0.9482435536 i$, and the numerical values obtained from Padé approximants are $\omega_3=0.2510825124+0.7047367862 i$ and $\omega_4= 0.2105327821+0.9461031269 i$. Compared with results in Table~\ref{tab:QNM_Taylor+Pade}, we can see that the QNM frequencies from Padé approximants are also improved by the homotopy deformation.

\textit{Conclusion---}In this work, inspired by Mashhoon's traditional method~\cite{Mashhoon1984}, we propose a new (numerical) method to obtain QNM frequencies from bound state spectra. 
In this new method, a single real parameter $\alpha$ is introduced via a coordinate scaling in the inverted potential well, and after calculating the bound state eigenvalues $E_n(\alpha)$, the QNM frequencies $\omega_n$ can be obtained by analytic continuation in $\alpha$ from  $E_n(\alpha)$.
In the traditional method, for each model, one need to find a corresponding transformation of the model parameters satisfying certain conditions. In contrast, 
the new method provides a unified and consistent treatment for different models and does not involve manipulation of the model parameters.  
As pointed in Refs.~\cite{Volkel2022,Volkel2025}, the numerical calculation of bound states is usually more stable than the direct computation of the corresponding QNMs, this new method is also interesting from a technical point of view.

As an application, we numerically compute various QNM frequencies for the SchBH using the exact Regge–Wheeler potential for the first time. 
Although similar numerical calculations with traditional method were pioneered by Völkel in Ref.~\cite{Volkel2022}, approximate potentials were used in that work. 
The new method allows direct application to the exact black hole potential without invoking approximate potentials.

The method yields QNM frequencies of high accuracy for low-lying modes of scalar, electromagnetic, and gravitational perturbations of the SchBH,
but fails for higher overtone modes. 
The analytic continuation in the new method is a single-complex-variable problem, and the numerical investigation of the analytic structure of $E_n(\alpha)$ is tractable via Padé approximants. A Padé analysis of these approximants identifies the approximate positions of the poles of $E_n(\alpha)$, which explains the failure to obtain higher overtones
QNM frequencies from bound state energies by analytic continuation. Then, a homotopy deformation scheme to the inverted potential is proposed, which can shift the obstructing singularities and allow computing reliable additional overtones.

In the main text, we just discuss the $\ell=2$ gravitational QNM frequencies for overtones $n\leq 4$. When $n>4$, it is found that the obstructing singularities of $E_n(\alpha)$ can no longer be pushed outward effectively, and the analytic continuation fails to produce reliable QNM frequencies.
An interesting further direction is to explore whether there is a conformal mapping in the complex $\alpha$-plane which could systematically shift the obstructing poles outside the analytic continuation circles. 
A straightforward direction is to apply the new method to calculate the QNM frequencies of other spherically symmetric black holes. 
It will also be interesting to explore similar single-parameter bound state method for Kerr and other rotating black holes admitting separation of variables.


\vspace{6pt} 
\textit{Acknowledgments---}This is a preliminary preprint; we welcome any comments to help improve this work. 

  \newpage

\bibliography{ref}


 \widetext 
\appendix

 \section{Numerical computation of bound state  eigenvalues}\label{sec:Appendix A}

We compute the bound state  eigenvalues $E_n(\alpha)$ using the shooting method\cite{Press2007}. For a trial energy $E<0$, we integrate Eq.~\eqref{eq:master} inward from two distant boundaries toward an intermediate matching point using an ninth-order Runge–Kutta (RK9) integrator, imposing the exponentially decaying boundary conditions:
\begin{equation}\label{boundary}
   \Psi_{\pm}(x)\sim e^{\mp \sqrt{-E }x},\,\,\,x\,\,\xrightarrow{} \pm\infty.
\end{equation}
A trial energy $E$ is an eigenvalue if the Wronskian of the two integrated solutions vanishes at the matching point, and the Wronskian is defined as:
\begin{equation}\label{eq:Wronskian}
    W(\Psi_- ,\Psi_+)(x,E)=\Psi_- \frac{d\Psi_+}{dx}
 -\frac{d\Psi_-}{dx}\Psi_+.
\end{equation}
Since the Wronskian can be regarded as a function of $E$, the eigenvalues $E_n$ can be found by computing the roots of the Wronskian. This one-dimensional root-finding problem can be solved numerically using the bisection method or the secant method.

 \section{Analytic continuation via Padé approximants}\label{sec:Appendix B}

The Padé approximant is a rational function approximation that exhibits higher accuracy and faster convergence than a Taylor series of the same order. It can naturally capture the poles of a function, extend its valid range via analytic continuation, and effectively avoid the Runge oscillations in high-order fitting. 

Using the bound state eigenvalues, we perform the analytic continuation from the expansion center $\alpha_0$ to the target point $\alpha=-i$ via Padé approximants. We first fit the discrete data set of $k$ points $\{E_n(\alpha_j)\}$ to a polynomial of degree $k-2$ using the least squares method:
\begin{equation}
E_n(\alpha) = \sum_{j=0}^{k-2} c_j (\alpha - \alpha_0)^j,
\label{eq:polyfit}
\end{equation}
where one point is reserved for Leave-One-Out Cross-Validation (LOOCV), reducing the maximum polynomial degree to $k-2$. From the coefficients $c_j$ of this polynomial, we then construct Padé approximants of the general form:
\begin{equation}
R_{[M/N]}(\alpha) = \frac{P_M(\alpha)}{Q_N(\alpha)} = \frac{\sum_{i=0}^{M} a_i (\alpha-\alpha_0)^i}{1 + \sum_{j=1}^{N} b_j (\alpha-\alpha_0)^j},
\label{eq:pade}
\end{equation}
where $M$ and $N$ are the degrees of the numerator and denominator polynomials, respectively, and satisfy $M+N\leq k-2$.  The coefficients $\{a_i,b_j\}$ are obtained by solving the linear system that results from matching the power-series coefficients of $R_{[M/N]}(\alpha)$ with those of $E_n(\alpha)$ up to the order $M+N$; only the coefficients $c_0,…,c_{M+N}$ of $E_n(\alpha)$ are involved in the construction. 

In this work, we only consider the diagonal ($[M/M]$), subdiagonal ($[M/M+1]$), and superdiagonal ($[M+1/M]$) Padé approximants. The optimal order $m$ is selected via LOOCV to mitigate overfitting and guarantee the reliability of results from analytic continuation. The QNM frequencies are then derived via the relation $\omega_n = \sqrt{-R_{[M/N]}(-i)}$, where $R_{[M/N]}(-i)$ is evaluated using the optimal order.

\section{QNM frequencies for scalar and electromagnetic perturbations}
\label{sec:Appendix D}

Tables~\ref{tab:QNM_scalar} and~\ref{tab:QNM_EM} present QNM frequencies for scalar  ($s=0$)  and electromagnetic ($s=1$) perturbations of SchBH, obtained with the new method incorporating the Padé  approximants. Reference values from Leaver’s continued fraction method are also provided. The results are of high accuracy for $n\leq\ell$, and the numerical accuracy gradually degrades as the overtone number $n$ 
increases. These characteristics are similar to that of the gravitational perturbations discussed in the main text.

\begin{table}[htbp]
\caption{QNM frequencies for scalar perturbations ($s=0$) of SchBH.}
 \begin{ruledtabular}
 \footnotesize  
 \begin{tabular}{c c c c c c}
$\ell$ & $n$ & $\omega_{\rm \text{Pad\'e}}$ & $\omega_{\rm Leaver}$&$\delta_{\rm Re}$& $\delta_{\rm Im}$\\
\midrule
$0$& $0$ & $0.110455263826298+0.104895448120225i$& $0.110454939080425+0.104895717086885i$
& $2.9\times10^{-06}$&$2.6\times10^{-06}$\\
\midrule
$1$& $0$ & $0.292936133271606+0.097659988916406i$& $0.292936133267283 + 0.097659988913578i$
& $1.5\times10^{-11}$&$2.9\times10^{-11}$\\
  & $1$ &$0.264448644859410+0.306257389941812i$&$ 0.264448650604837+ 0.306257391559060i$& $2.2\times10^{-08}$&$5.3\times10^{-09}$\\
\midrule
$2$& $0$ & $0.483643872210710+0.096758775978288i$& $0.483643872210713 + 0.096758775978288i$
& $6.2\times10^{-15}$&$3.2\times10^{-15}$\\
  & $1$ & $0.463850579020222+0.295603936988600i$& $0.463850579019766 + 0.295603936987963i$
& $9.8\times10^{-13}$&$2.2\times10^{-12}$\\
  & $2$ & $0.430544073921966+0.508558364630691i$&$ 0.430544054376576 + 0.508558402154275i$
& $4.5\times10^{-08}$&$7.4\times10^{-08}$\\
 \end{tabular}
\end{ruledtabular}
\label{tab:QNM_scalar}
\end{table}

\begin{table}[htbp]
\caption{QNM frequencies for electromagnetic perturbations ($s=1$) of SchBH. }
 \begin{ruledtabular}
 \footnotesize  
 \begin{tabular}{c c c c c c}
$\ell$ & $n$ & $\omega_{\rm \text{Pad\'e}}$ & $\omega_{\rm Leaver}$&$\delta_{\rm Re}$& $\delta_{\rm Im}$\\
\midrule
$1$& $0$ & $0.248263264178682+0.092487717955451i$& $0.248263264178109 + 0.092487717952942i$
& $2.3\times10^{-12}$&$2.7\times10^{-11}$\\
  & $1$ & $0.214515437283184+0.293667663922264i$
& $0.214515419563606+ 0.293667645545729i$
& $8.3\times10^{-08}$&$6.3\times10^{-08}$\\
\midrule
$2$& $0$ & $0.457595511629852+0.095004425819475i$
& $0.457595511629853 + 0.095004425819472i$
& $2.2\times10^{-15}$&$3.7\times10^{-14}$\\
  & $1$ &$0.436542385747294+0.290710143117091i$
& $0.436542385750543 + 0.290710143120364i$
& $7.4\times10^{-12}$&$1.1\times10^{-11}$\\
  & $2$ & $0.401186709722653+0.501587349841500i$
& $0.401186733916441 + 0.501587346341654i$
& $6.0\times10^{-08}$&$7.0\times10^{-09}$\\
\midrule
$3$& $0$ & $0.656898670462478+0.095616217928335i$
& $0.656898670462478+0.095616217928335i$
& $4.3\times10^{-17}$&$2.2\times10^{-16}$\\
  & $1$ & $0.641737435968029+0.289728401728145i$
& $0.641737435967901 + 0.289728401728205i$
& $2.0\times10^{-13}$&$2.1\times10^{-13}$\\
  & $2$ & $0.613832026334886+0.492066258236375i$
& $0.613832026334250 + 0.492066258237909i$
& $1.0\times10^{-12}$&$2.1\times10^{-12}$\\
  & $3$ & $0.577918499877886+0.706330830669244i$
& $0.577918506078184 + 0.706330830287275i$
& $1.1\times10^{-08}$&$5.4\times10^{-10}$\\
 \end{tabular}
\end{ruledtabular}
\label{tab:QNM_EM}
\end{table}

\end{document}